# Adverse Effects of Polymer Coating on Heat Transport at Solid-Liquid Interface

Shenghong Ju[1], Bruno Palpant[1], and Yann Chalopin[2,*]

[1]Laboratoire de Photonique Quantique et Moléculaire, CentraleSupelec, Ecole Normale Supérieure Paris-Saclay, Université Paris Saclay, CNRS, UMR 8537, Grande Voie des vignes F-92295 Châtenay-Malabry, France

[2]Laboratoire Energétique Macroscopique et Moléculaire, CentraleSupelec, Université Paris Saclay, CNRS, UPR 288, Grande Voie des vignes F-92295 Châtenay-Malabry, France

*Corresponding email: yann.chalopin@cnrs.fr

**ABSTRACT**

The ability of metallic nanoparticles to supply heat to a liquid environment under exposure to an external optical field has attracted growing interest for biomedical applications. Controlling the thermal transport properties at a solid-liquid interface then appears to be particularly relevant. In this work, we address the thermal transport between water and a gold surface coated by a polymer layer. Using molecular dynamics simulations, we demonstrate that increasing the polymer density displaces the domain resisting to the heat flow, while it doesn't affect the final amount of thermal energy released in the liquid. This unexpected behavior results from a trade-off established by the increasing polymer density which couples more efficiently with the solid but initiates a counterbalancing resistance with the liquid.

## I. INTRODUCTION

Thermal transport at solid-liquid interface [1-9] has been extensively investigated in various areas of science, including basic research such as condensed-matter physics, nanoscale heat transfers, or transport in porous media. In addition, this topic is of practical importance to the development of strategies that allow to increase the performances of engineering systems. Moreover, recent progress that have enabled for tailoring materials at the nanoscale have opened the window on a wide range of applications in biotechnologies and engineering [10-14]. In any of these cases, understanding the heat exchanges at the interface between matter and a surrounding liquid are of fundamental importance. Due to their unique optical properties, gold nanoparticles can supply thermal energy at the nanoscale with the mean of an external light excitation [15-23]. Depending on the targeted applications, these nanoparticles usually implies a coating with different classes of molecules [24-30]. Understanding the resistance to the heat flow of a solid surface coated with complex molecular structure in thermal interaction with a liquid is crucial, although it remains a challenging issue due to the nature of the thermal interfaces involved. On one hand, heat exchanges between the solid part and an array of polymer involves the vibrational spectra of the contact atoms. The corresponding resistive process is driven by the mismatch of the phonon modes on both sides [31,32]. On the other hand, a second class of interface (formed by the metal or the polymer and the liquid) implicates momentum exchange through molecular collisions. Characterizing and estimating how these two contributions are balanced is the main line of this work. Hence, the discussion goes as follow: having a hot solid surface coated with polymer, how does it affect the heating of the liquid? In addition, how does this heating depends on the polymer density? To answer these questions, we have quantitatively calculated the thermal resistance of such a system by using an atomistic approach based on molecular dynamics simulations, in which we reproduce a numerical experiment where the system is constantly maintained out-of-equilibrium by a temperature gradient. We demonstrate that unexpectedly the global thermal resistance of such a complex metal-polymer-water interface remains constant whatever the polymer density. The analysis reveals that this surprising effect results from a trade-off between two interfacial conductances evolving oppositely with the grafting concentration.

## II. MODELS AND SIMULATIONS

The polymer used in this study is the Jeffamine M600 [33,34] which is a block copolymer constituted of assembled ethylene oxide (EO) and propylene oxide (PO) groups. It is a thermosensitive polymer that exhibit a phase transition in water from hydrophobic to hydrophilic above a certain temperature. Such polymers grafted on plasmonic nanostructures enable then to achieve promising optically-driven functionalities through photothermal conversion [26,29,35-38]. These amino-terminated polymer is modified by dihydrolipoic acid (DHLA). This excellent anchor group provides a high stability of the polymer ligand on the gold surface compared to monothiol group [39,40]. In addition, the Jeffamine family of polymer offers interesting features that can be tuned by adjusting the ratio between EO and PO groups. However, this aspect remains out of the scope of this study in which the temperature range will always remain lower than the polymer transition temperature. In this study, the geometry of the nanoparticle is not accounted for, i.e., we have considered the limit where the surface curvature tends to zero. Figure 1 presents the system of interest. The corresponding atomic configuration is particularly suited to perform non-equilibrium molecular dynamics (NEMD) simulation [41-43]. It consists of three parts: the gold crystal (left and right), and in between the water region containing polymer chains grafted on one of the two metallic surfaces. The cross-section of the system is 4.2 nm by 3.6 nm. The thickness of the gold slabs is 4.1 nm, followed by the liquid (from 7.2 nm to 10 nm depending on the grafting density). In order to ensure the numerical stability of the system, the two metallic layers include fixed boundary conditions at the extremities (indeed, in a molecular dynamics approach, controlling the temperature in a solid is much easier than in a liquid). Thermal energy is injected and subtracted at the two sides, respectively, by rescaling the velocities of the atoms with two thermostats set at a high $T_1$ and a low $T_2$ temperatures. This temperature difference maintains the system out of equilibrium, and a heat flux $J$ is generated. It becomes then possible to estimate the different values of the conductances $G$ by tracking the various temperature drops $\Delta T$ using

$$G = \frac{J}{A\Delta T}, \quad (1)$$

where $A$ is the cross section area perpendicular to the heat flux. In order to unveil the effect of the polymer on heat transfer, various densities ranging from 0 to 3.2 chains/nm$^2$ are considered (as shown in Fig. 2).

The interactions (the force fields) between the polymer atoms were accounted for by the DREIDING model [44], which is widely employed to describe the structures and the dynamics of organic and biological molecules. The MEAM potential [45] has been set for the gold lattice. The TIP3P potential [46,47] models the water molecules. Long-range coulomb interactions are calculated using the particle-particle particle-mesh solver (PPPM) [48,49]. The total number of atoms $N$, the temperature $T$ and the atmosphere pressure $P = 1$ atm are maintained constant (NPT ensemble) during an equilibrium procedure for a time of 1 ns, using the leapfrog integration algorithm. The numerical time step has been set to 1 fs. The temperature differences are next established in the microcanonical ensemble for 4 ns to reach the steady state. From this basis, the temperature profile is extracted from the atomic velocities all the way down between the two thermostats.

## III. RESULTS AND DISCUSSION

It has been recently demonstrated that the phonon transmission function that characterizes any interfacial resistance effect is related to the cross-correlation between the local density of states [31] of the media in contact. To define which vibrational frequencies are involved at each interface, we have calculated the vibrational density of states, as shown in Fig. 3. This is done by considering the atomic velocity fluctuations $v_i$ of each atom $i$ belonging to any of the three systems (polymer, gold or water),

$$g(\omega) = \frac{1}{k_B T} \sum_i m_i \int_{-\infty}^{+\infty}\int_{-\infty}^{+\infty} \vec{v}_i(t) \cdot \vec{v}_i(t+\tau) e^{i\omega t} dt d\tau. \tag{2}$$

An analysis of the structure of the density of states reveals that the transport of heat occurs with a vibrational energy ranging from 1 to 10 THz. This frequency band corresponds to a domain where the 3 systems (polymer, Au and water) exhibit a common basis of vibrational modes allowing the heat energy exchange. Considering that the lowest frequencies of the solid interface are associated to the higher group velocities, it is very likely that this low frequency overlap plays the main role in the heat flux. Figure 2 presents the temperature distribution profiles obtained for several concentrations ranging from 0 to 3.2 chains/nm$^2$. These temperature profiles are correlated with the local atomic density for the three species (gold, water and polymer). We have indicated on the same figure the various interface resistances that arise when increasing the polymer concentration. The first one corresponds to that between gold and water $R_{g-w}$ as well as a resistance between the gold and the polymer chains $R_{g-p}$. Another interface resistance between the polymer and water $R_{p-w}$ also appears. It can be seen that as the polymer concentration increases, most of the water molecules tend to be rejected from the metal surface. Consequently, the thermal resistance at the interface between the gold and the polymer progressively replaces that between the gold and the liquid. Concomitantly, an additional resistance between the polymer and the water arises. At the highest concentration of 3.2 chains/nm$^2$ (which is closed to what can be achieved experimentally) one can notice that the release of heat from the gold surface to the water occurs through two dominant interfacial resistances $R_{g-p}$ and $R_{p-w}$. It can be seen than the conductance of the polymer is quite high suggesting that ballistic transport occurs in these layers. From Fig. 2, we can conclude qualitatively that increasing the polymer density 1) decreases the first interface resistance between the gold surface and the hybrid water/polymer system; 2) the conductance within the region containing polymer increases; 3) an additional interface resistance between the polymer and the water progressively raises. From this consideration, a simple question arises: What is the hierarchy between theses resistive processes and how does it affect the heating of the liquid?

To address this question, calculations have been conducted for different temperatures compatible with biological applications (i.e., from 300 K to 360 K). The case of the highest density of polymer (3.2 chains/nm$^2$) has been considered. Figure 4 presents quantitatively the hierarchy obtained between the gold-polymer and the polymer-water interface resistances compared to the case of the pure gold-water interface. It can be observed that the gold-water interface resistance is predominant and roughly three times larger than the interface resistance between the polymer and the water which presents the smallest contribution. A value of 108 MW/m$^2$K at 300 K has been obtained which is very similar to previously reported 100-300 MW/m$^2$K by both experiment [50,51] and simulation [52]. The ranking between the various interfacial resistances remains stable and barely does not depend on temperature.

From this analysis of the heat transport at the highest possible polymer concentration, a simple question arises: how can we explain and model the thermal conductance between the solid and the liquid at lower polymer concentrations? In what follows, we derive a model revealing that the thermal resistance $R_{hybrid}$ between the gold and the water/polymer region can be viewed as a parallel process driven by two distinct resistances: that between gold and water in parallel with that between the polymer and the water.

For this purpose, we divided the interface into several small unit surface, in which only one chain of polymer can be bonded. In this way, the hybrid interface was composed of $N_{(gw)}$ of gold-water units and $N_{(gp)}$ of gold-polymer units. This allows us to define the hybrid resistance of the coated gold surface in contact with water as a dual contribution coming from the gold-water unit resistance $R_{(gw)}$ plus the gold-polymer unit resistance $R_{(gp)}$. This hybrid resistance is thus define with the parallel resistance rule as,

$$\frac{1}{R_{hybrid}} = \frac{N_{(gw)}}{R_{(gw)}} + \frac{N_{(gp)}}{R_{(gp)}}. \tag{3}$$

The hybrid interface resistance predicted at different polymer density is shown on Fig. 5 and compared to direct molecular dynamics simulations to validate the model. The two approaches match with a satisfactory agreement that confirms the hypothesis of an interface resistance involving mixed species (i.e., polymer and water corresponding to parallel thermal channels).

These considerations finally allow us to estimate the total thermal resistance from the gold surface to the water and that, as a function of the grafted polymer concentration. The total thermal resistance is decomposed as the sum of three contributions: the first hybrid resistance $R_{hybrid}$ between gold and the polymer-water mixture, the resistance in the polymer-water mixture layer region called $R_{pw}$ in addition to that between the end of the polymer chains and water $R_{p-w}$. Figure 6 presents these various contributions as well as it indicates the corresponding total resistance for the different concentrations (the associated temperature profiles remain those of Fig. 2). What can be concluded from this is that the increase of polymer density leads to a decrease of the hybrid resistance. Concomitantly, the interface between polymer and water gets formed and the corresponding resistance increases with the grafting concentration. For the layer where the polymer and the water coexist, the thermal resistance decreases due to the increasing contribution of the polymer in which heat propagates ballistically (see Fig. 2(e) for instance) where almost no temperature drop is observed for the case where the water has been completely excluded from the polymer region. Consequently, increasing the polymer density allows the heat to escape more efficiently from the gold surface. Then thermal energy propagates with almost no losses within the polymer chains. However, as thermal energy flows to the termination of the latter, the transport of heat is refrained by an important resistance with the water. This resistance dramatically diminishes the release of thermal energy to the liquid. A quantitative analysis of the interplay between these contributions allows us to conclude that grafting the surface of a metal particle with a polymer does not influence the amount of heat relaxed in the liquid whatever the polymer density. In fact, increasing the density produces a trade-off that displaces the location where the predominant thermal resistance occurs but it does not allow to significantly achieve any increase of the temperature in the liquid. Consequently, at a distance far from the polymer, the elevation of the liquid temperature is an invariant of the concentration of polymer grafted on the metal surface.

## IV. CONCLUSIONS

In summary, we have studied the mechanisms of heat release from a solid surface to a liquid from an atomistic modeling. To address this problem we have considered a gold surface coated with a polymer in contact with water. The effect of the polymer grafting has been investigated in details. As the contact resistance between the polymer and the gold surface is much lower than that between the gold and the water, it could intuitively indicate that any increase the polymer density may eventually lead to optimize the heating process of the liquid. In fact, we demonstrate that despite these polymer chains (here Jeffamine) are extremely good thermal conductors, the thermal coupling with water molecules is rather inefficient. Consequently, as the polymer grafting density increases, the predominant resistance effect gets progressively displaced from the metal surface to the region where the polymer endings interact with the water molecules. A quantitative estimation of the trade-off allows us to conclude that the total thermal resistance remains independent of the polymer concentration. This important result provides practical strategy for the design of nano-sources of heat for biological applications as well as it allows us to define a framework to unveil in detail the various mechanisms occurring at complex interfaces involving fluid, liquid and soft materials. Beyond, to improve the heat transfer from a solid to the water, an interesting strategy would be to precisely engineer the termination of the grafted polymer in contact with water in view to lower its contact resistance.

This work was financially supported by the French National Research Agency (ANR) under the grant ANR-13-BS10-0008. The authors acknowledge the useful discussions with Nicolas Sanson, Simona Cristina Laza and Jordane Soussi. The calculations in this work were performed using IGLOO clusters of CentraleSupelec.

**Figures and captions**

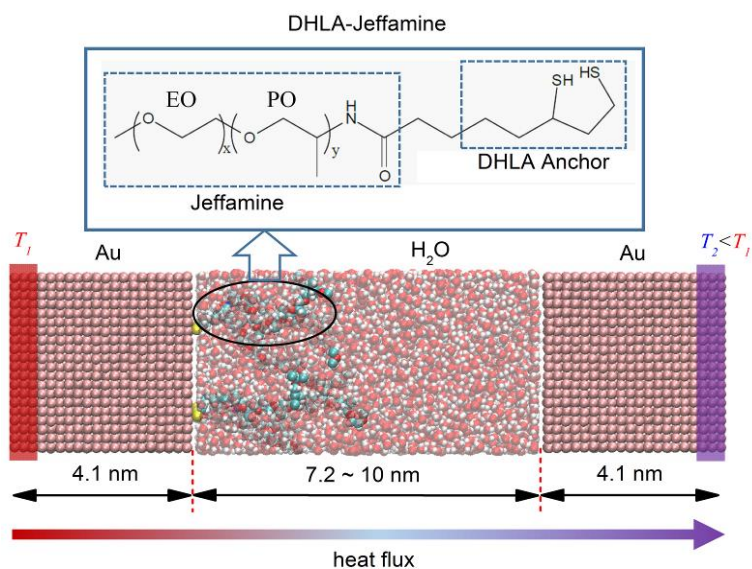

**Fig. 1** Scheme of the total system considered for the NEMD simulations. The cross-section size of the system is 4.2 nm × 3.6 nm. The coated plymers on gold surface is composed of two parts: dihydrolipoic acid (DHLA) anchor and Jeffamine polymer. The total number of water moleculars keeps constant for different grafted density of polymers.

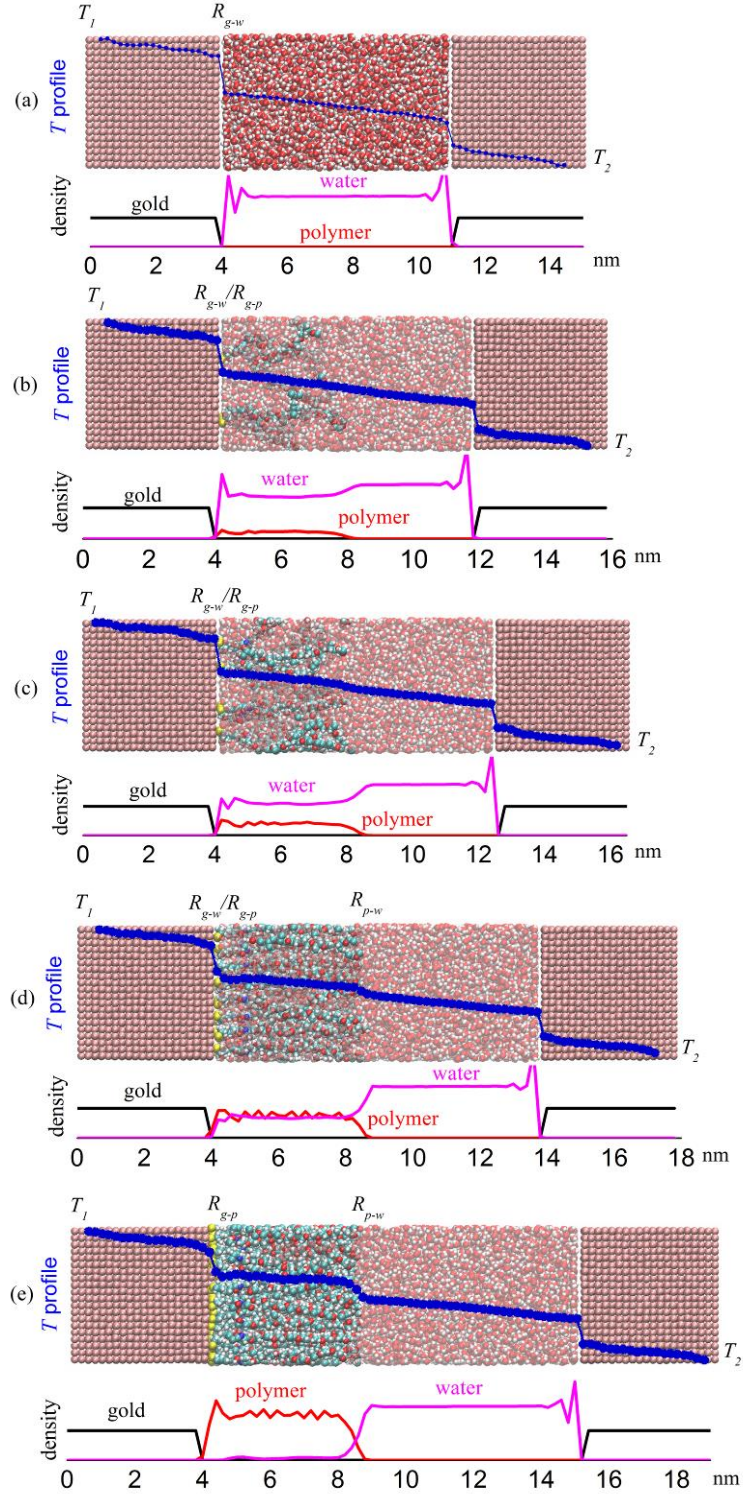

**Fig. 2** Effect of the grafting density on the atomic density and the temperature profile along the gold-polymer-water system: (a) no polymers, (b) grafting density of 0.4 chain/nm$^2$, (c) 0.8 chain/nm$^2$, (d) 1.6 chain/nm$^2$, (e) 3.2 chain/nm$^2$.

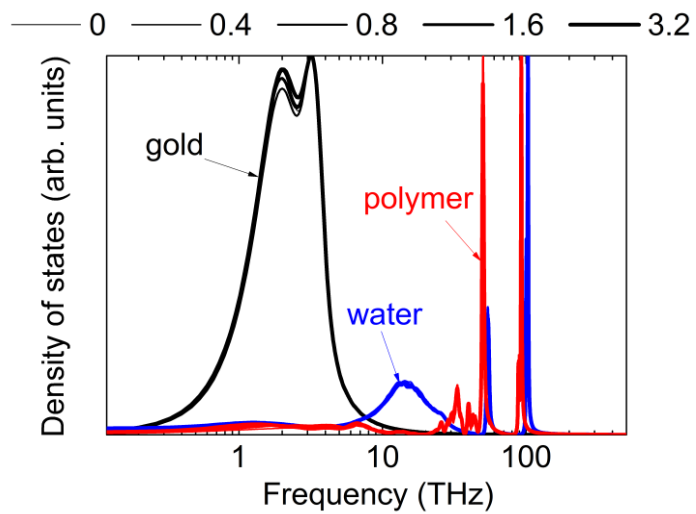

**Fig. 3** Vibrational density of states at different polymer grafting densities (0 to 3.2 chains/nm$^2$). Vibrational density of states calculated using Eq. 2 and decomposed for the water, the gold surface in contact with the polymer and the polymer atoms.

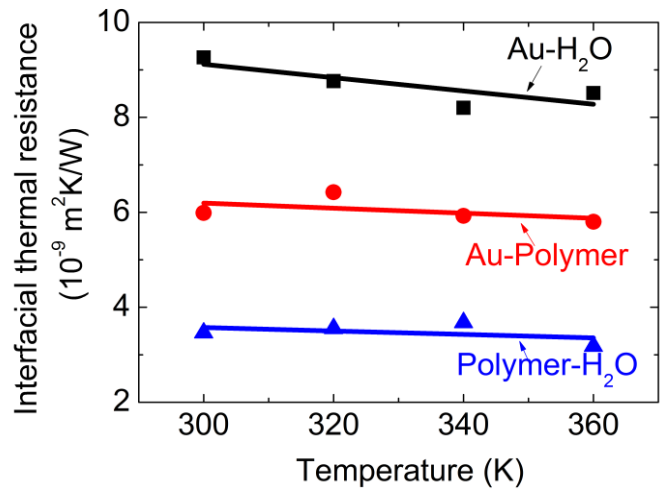

**Fig. 4** Pure interfacial thermal resistances at different temperatures in a hybrid gold-polymer-water system for a concentration of 3.2 chains/nm$^2$. The rectangular, circular and triangular dots show the ITR values between gold and water $R_{g-w}$, gold and polymer $R_{g-p}$, and polymer and water $R_{p-w}$, respectively.

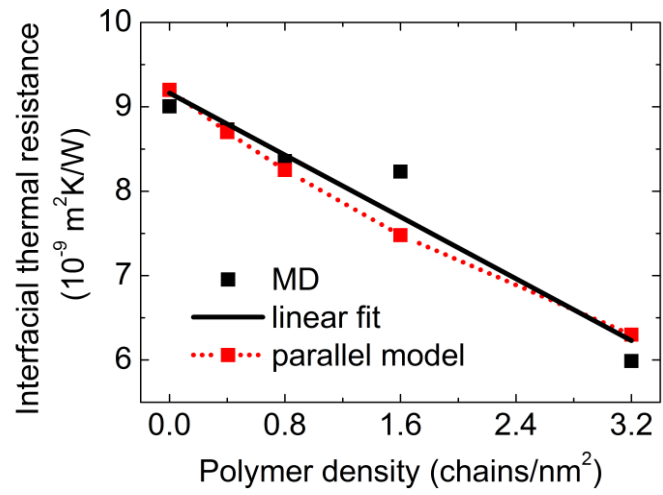

**Fig. 5** Comparison of hybrid resistance between the gold:water+polymer region predicted by parallel model and direct molecular dynamics simulations.

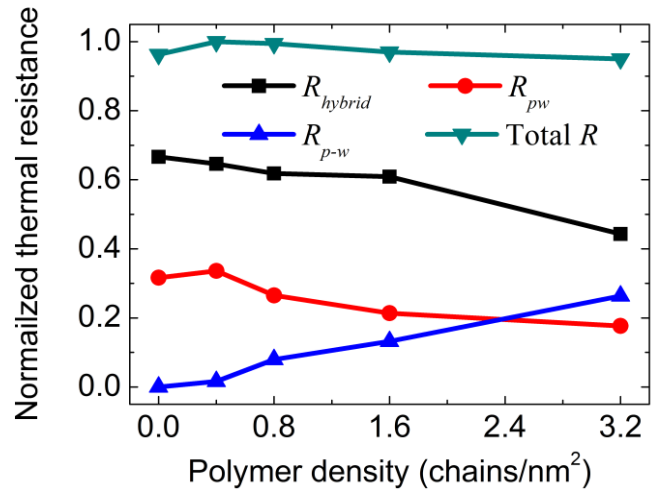

**Fig. 6** Decomposition of the total thermal resistance between the gold surface and the liquid at different polymer grafting densities. The resistances are normalized by dividing the data by the maximum total resistance value found among all the polymer density cases.